\documentclass[aps,prl,reprint,twocolumn,10pt,showpacs,superscriptaddress,footinbib]{revtex4-1}
\usepackage{amsmath}
\usepackage{amsthm}
\usepackage{graphicx}
\usepackage{times}
\usepackage{amssymb}
\usepackage{hyperref}
\hypersetup{
colorlinks=true,
linkcolor=red,
citecolor=red,}
\usepackage{textcomp}
\usepackage{xcolor}
\usepackage{enumerate}
\usepackage[final]{pdfpages}
\usepackage{multirow}
\usepackage{tabularx}
\usepackage{dcolumn}
\usepackage[mathscr]{euscript}
\usepackage{mathtools}
\usepackage{needspace}
\usepackage[mathscr]{euscript}

\begin{document}
\title{Measurement-Device-Independent Approach to Entanglement Measures}
 \author{Farid Shahandeh}
 \email{Electronic address: f.shahandeh@uq.edu.au}
  \affiliation{Centre for Quantum Computation and Communication Technology, School of Mathematics and Physics, University of Queensland, Brisbane, Queensland 4072, Australia}
 \author{Michael J. W. Hall}
 \affiliation{Centre for Quantum Computation and Communication Technology, Centre for Quantum Dynamics, Griffith University, Brisbane, Queensland 4111, Australia}
 \author{Timothy C. Ralph}
 \affiliation{Centre for Quantum Computation and Communication Technology, School of Mathematics and Physics, University of Queensland, Brisbane, Queensland 4072, Australia}

\begin{abstract}
Within the context of semiquantum nonlocal games, the trust can be removed from the measurement devices in an entanglement-detection procedure. 
Here we show that a similar approach can be taken to quantify the amount of entanglement.
To be specific, first, we show that in this context a small subset of semiquantum nonlocal games is necessary and sufficient for entanglement detection in the LOCC paradigm.
Second, we prove that the maximum pay-off for these games is a universal measure of entanglement which is convex and continuous.
Third, we show that for the quantification of negative-partial-transpose entanglement, this subset can be further reduced down to a single arbitrary element.
Importantly, our measure is operationally accessible in a measurement-device-independent way by construction.
Finally, our approach is simply extended to quantify the entanglement within any partitioning of multipartite quantum states.
\end{abstract}

% \pacs{...}

\maketitle

\paragraph*{Introduction.---}

Entanglement is a valuable resource for practical as well as fundamental applications of quantum theory, ranging from quantum computation and communication to metrology~\cite{Nielsen,Horodecki2009,LIGO}.
There are two major challenges in understanding entanglement that stimulates this research.
It is extremely difficult to specify all the nonentangled bipartite or multipartite quantum states.
In fact, the problem is known to be NP-hard~\cite{Gurvits2003,Gharibian2010}.
Not surprisingly, the characterization of entangled states is an equally difficult task.
That is, to quantify the amount of entanglement within a quantum state.
The answer to this question is practically very important, because it tells us how well our protocols will perform using a given state~\cite{Furusawa98,Parker2000,Gross2009,Gross2010}.

\begin{figure}[h]
  \includegraphics[width=\columnwidth]{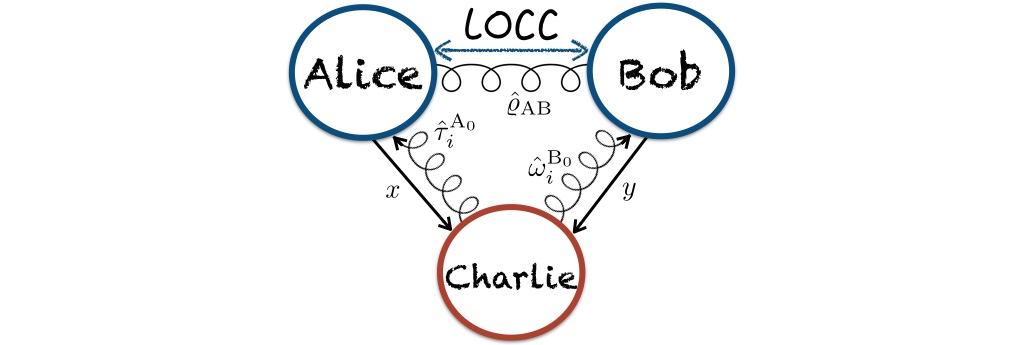}
  \caption{The scheme of a semiquantum nonlocal game.
  Charlie asks the players quantum questions while the players return classical answers.
  The shared state between the players helps them to obtain a maximum pay-off in the game.
  Here, we allow the players to access LOCC and introduce a measurement-device-independent measure of entanglement for the shared state.
  }\label{protocol}
\end{figure}

Recent work by Buscemi~\cite{Buscemi2012} has introduced a new way to think about entanglement detection~\cite{Branciard2013,Xu2014,Nawareg2015,Verbanis2016}.
The idea is to map the problem onto a modified class of nonlocal games, called semiquantum nonlocal games (SQNLGs).
In any such game, two players (Alice and Bob) share a possibly entangled state.
A referee (Charlie) starts asking them quantum questions by sending quantum states and receiving classical answers---the outcomes of some local measurements (see Fig~\ref{protocol}).
He then evaluates a {\it reward} function from coincidence statistics and pays the players accordingly.
Confined to not communicate during the game, known as the local operations and shared randomness (LOSR) paradigm, all separable states deliver an equal {\it pay-off} (maximum average reward) in any specific SQNLG.
Importantly, for every entangled quantum state one can always find a SQNLG which can deliver a larger pay-off than any separable state.
This mapping allows one to merely rely on the coincidence statistics of measurement outcomes, rather than their specific expectation values, to violate an entanglement witnessing inequality~\cite{Branciard2013}; a property that was believed to be specific to Bell nonlocality scenarios.
Consequently, researchers interpreted Buscemi's results as a clever way to remove the trust from measurement devices in an entanglement witnessing procedure, since any entanglement witness (EW) can be recast as a SQNLG~\cite{Branciard2013,Cavalcanti2013}.

Focusing on the second challenge, a first level of hardness is that the quantification of entanglement using almost any entanglement measure, e.g., entanglement of formation~\cite{Bennett96}, negativity~\cite{Zyczkowski98,Vidal2002}, or random robustness~\cite{Vidal99}, requires estimating a large number of density matrix elements; a task which is difficult to perform on bipartite and multipartite quantum states.
While this difficulty can be partially circumvented by making use of EWs when lower bounds on the entanglement are desired~\cite{Brandao2005,Eisert2007,Sperling2011,Lee2012,Shahandeh2013,Shahandeh2014}, errors and misalignments of the measurement devices can still lead to the incorrect estimations of the quantities and thus, erroneous conclusions.

In this Letter, inspired by Buscemi's approach, we consider SQNLGs in the paradigm of local operations and classical communication (LOCC).
We show that a small subset of games which we call extremal semiquantum witnessing games (ESQWGs) are both necessary and sufficient for the full characterization of entangled states.
We then focus on the entanglement of negative-partial-transpose (NPT) states, as the necessary ingredient for distillability~\cite{Horodecki99}.
We present the answer to the practical appeal for a measurement-device-independent (MDI) NPT-entanglement measure by proving that NPT entanglement can be quantified by a referee in {\it a single arbitrary} ESQWG.
The main result of the present Letter is thus to introduce an operationally accessible measure (for both general and NPT) entanglement that is convex, universal, and most importantly MDI by construction.
Furthermore, we extend our measure to quantify the entanglement in all possible partitions of multipartite quantum states.

%======================================================
%======================================================
%				From SQNLGs to SQWGs

\paragraph*{From SQNLGs to SQWGs.---}

Let us start by describing SQNLGs more rigorously.
A SQNLG is a collaborative game, denoted here by $\mathsf{G}_{\rm sq}$, in which Alice and Bob share a quantum state $\hat{\varrho}_{\rm AB}$.
Charlie prepares a set of quantum questions $\{\hat{\tau}^{{\rm A}_0}_i\}$ and $\{\hat{\omega}^{{\rm B}_0}_i\}$ with probability $\{p_i\}$ and sends them to Alice and Bob, respectively.
The players respond to each question classically from two sets of labels $\{x\}_{\rm A}$ and $\{y\}_{\rm B}$.
Before the game starts, given the LOSR paradigm, they can agree on a best strategy to win the game, however, during the game they are no longer allowed to communicate.
For each question $i$, Charlie evaluates a reward corresponding to the answers $x$ and $y$ according to the function $\wp(x,y|i)$.
The average reward of the game is then given by 
\begin{equation}\label{AVpayoff}
\overline{\wp}(\hat{\varrho}_{\rm AB};\hat{P}^{\tilde{\rm A}};\hat{Q}^{\tilde{\rm B}};\mathsf{G}_{\rm sq})=\sum_{i,x,y} p_i \wp(x,y|i) \mu(\hat{P}^{\tilde{\rm A}}_x,\hat{Q}^{\tilde{\rm B}}_y|i,\hat{\varrho}_{\rm AB}),
\end{equation}
in which ${\tilde{\rm A}}\equiv {{\rm AA}_0}$, ${\tilde{\rm B}}\equiv{{\rm B}_0{\rm B}}$, and the joint probability distribution $\mu(\hat{P}^{\tilde{\rm A}}_x,\hat{Q}^{\tilde{\rm B}}_y|i,\hat{\varrho}_{\rm AB})$ is
\begin{equation}\label{PrMeasure1}
{\rm Tr}(\hat{P}^{\tilde{\rm A}}_x \otimes \hat{Q}^{\tilde{\rm B}}_y)(\hat{\tau}^{{\rm A}_0}_i \otimes \hat{\varrho}_{\rm AB} \otimes \hat{\omega}^{{\rm B}_0}_i),
\end{equation}
where $\hat{P}^{\tilde{\rm A}}_x \in \mathcal{M}_{\tilde{\rm A}}$ and $\hat{Q}^{\tilde{\rm B}}_y\in\mathcal{M}_{\tilde{\rm B}}$ are local effects (POVM elements) of the players.
They win or lose some value if the average reward is positive or negative, respectively.

The players' goal is of course to maximize the average amount they can obtain in a game.
Let us call the maximum average reward the {\it pay-off} value~\cite{foot1} and denote it by
\begin{equation}\label{measure1}
\wp^\star(\hat{\varrho}_{\rm AB};\mathsf{G}_{\rm sq}) = \max_{\hat{P}^{\tilde{\rm A}},\hat{Q}^{\tilde{\rm B}}} \overline{\wp}(\hat{\varrho}_{\rm AB};\hat{P}^{\tilde{\rm A}};\hat{Q}^{\tilde{\rm B}};\mathsf{G}_{\rm sq}).
\end{equation}
The main result of Buscemi~\cite{Buscemi2012}, relevant for entanglement detection, can be recast as follows. 
Given the set of all SQNLGs, $\mathcal{G}_{\rm sq}$, and the set of all separable states, $\mathcal{S}_{\rm sep}$, for any game $\mathsf{G}_{\rm sq} \in \mathcal{G}_{\rm sq}$ and for any two states $\hat{\varrho}_{\rm AB},\hat{\sigma}_{\rm AB} \in \mathcal{S}_{\rm sep}$, one has 
\begin{equation}\label{SepSQNLequal}
\wp^\star(\hat{\varrho}_{\rm AB};\mathsf{G}_{\rm sq}) = \wp^\star(\hat{\sigma}_{\rm AB};\mathsf{G}_{\rm sq}) = \wp^\star(\mathcal{S}_{\rm sep};\mathsf{G}_{\rm sq}).
\end{equation}
This simply reads as {\it all separable quantum states, at best, are equal in a SQNLG}.

\paragraph*{Criterion~1. (Buscemi).}
A quantum state $\hat{\varrho}_{\rm AB}$ is entangled if and only if there exists a SQNLG for which $\wp^\star(\hat{\varrho}_{\rm AB};\mathsf{G}_{\rm sq}) > \wp^\star(\mathcal{S}_{\rm sep};\mathsf{G}_{\rm sq})$.

It is relevant to ask whether one should search within the whole set $\mathcal{G}_{\rm sq}$ for a game violating the equality~\eqref{SepSQNLequal}?
The short answer is in negative~\cite{Branciard2013}.
Without loss of generality, we assume that the Hilbert spaces are finite dimensional, since entanglement can always be verified in finite dimensional subspaces~\cite{Sperling2009Fin}.
It is well-known that, by Hahn-Banach theorem, for any entangled state $\hat{\varrho}_{\rm AB}\notin\mathcal{S}_{\rm sep}$ there exist an EW $\hat{W}$ such that
\begin{equation}\label{Wineq}
{\rm Tr}\hat{W}\hat{\varrho}_{\rm AB} > 0,\text{~and~}\forall\hat{\sigma}_{\rm AB}\in\mathcal{S}_{\rm sep},~{\rm Tr}\hat{W}\hat{\sigma}_{\rm AB} \leqslant 0.
\end{equation}
Note that, here, for the sake of consistency, we have changed the sign of the usual convention.
Moreover, we set ${\rm Tr}\hat{W}=-D$, with $D=\min\{d_A,d_B\}$ being the minimum dimensionality of Alice and Bob's subsystems, to compare different EWs, and we note that such a normalization is always possible~\cite{Lewen2000}.
Now, every EW can be transformed into a SQNLG as follows.
Charlie decomposes the witness in terms of product states as $\hat{W}=\sum_i \beta_i \hat{\tau}^{{\rm A}_0\mathsf{T}}_i \otimes \hat{\omega}^{{\rm B}_0\mathsf{T}}_i$, with $\mathsf{T}$ denoting the transposition operation and $\beta_i\in\mathbb{R}$, and defines a SQNLG via
\begin{equation*}
\hat{W} \leftrightarrow \mathsf{W}_{\rm sq} \Leftrightarrow
\wp(x,y|i)= 	\left(\frac{\beta_i}{p_i}\right) \delta_{1,x}\delta_{1,y}.
\end{equation*}
It is important to note that this correspondence is not unique.
Now, we can rewrite Eq.~\eqref{measure1} as
\begin{equation}\label{SQWG}
\wp^\star(\hat{\varrho}_{\rm AB};\mathsf{W}_{\rm sq}) = \max_{\hat{P}^{\tilde{\rm A}},\hat{Q}^{\tilde{\rm B}}} {\rm Tr}(\hat{P}^{\tilde{\rm A}}_1 \otimes \hat{Q}^{\tilde{\rm B}}_1)(\hat{W}\otimes\hat{\varrho}_{\rm AB})\\
\end{equation}
We call any such a game a semiquantum witnessing game (SQWG) and denote the set of all such games by $\mathcal{W}_{\rm sq}$.
We may point out here that, the decomposition of the EW can be generalised to be in a nonseparable basis.
Branciard~{\it et al}~\cite{Branciard2013} showed that the set $\mathcal{W}_{\rm sq}$ is, indeed, necessary and sufficient for verifying the entanglement of a state $\hat{\varrho}_{\rm AB}$ shared by the players.
That is, $\wp^\star(\mathcal{S}_{\rm sep};\mathsf{W}_{\rm sq})\leqslant 0$, and for any entangled state $\hat{\varrho}_{\rm AB}\notin\mathcal{S}_{\rm sep}$ there exist a SQWG such that $\wp^\star(\hat{\varrho}_{\rm AB};\mathsf{W}_{\rm sq})> 0$.

%======================================================
%======================================================
%				Extremal SQWGs

\paragraph*{Extremal SQWGs.---}

In general, every EW is (with our sign convention) a member of the compact convex set of normalized block-negative operators (defined as operators with negative expectation values in all pure product states).
An extremal EW (EEW) $\hat{W}_{\rm e}$ is the one that cannot be written as a convex combination of any two other block-negative operators, and hence, there exists a pure product state $|a,b\rangle\in\mathcal{S}_{\rm sep}$ such that $\langle a,b|\hat{W}_{\rm e}|a,b\rangle = 0$.
We now introduce the set of {\it extremal semiquantum witnessing games} (ESQWGs), $\mathcal{W}^{\rm e}_{\rm sq}\subset\mathcal{W}_{\rm sq}$, which correspond to EEWs.
This class of games is necessary and sufficient for entanglement detection, since for every entangled state there exists an EEW which detects it~\cite{Chruscinski2014,Shultz2016}.
A very important corollary thus follows.
For any $\mathsf{W}^{\rm e}_{\rm sq}\in\mathcal{W}^{\rm e}_{\rm sq}$, we have that~\cite{Supp}
\begin{equation}\label{maxpayoffSep}
\wp^\star(\mathcal{S}_{\rm sep};\mathsf{W}^{\rm e}_{\rm sq}) = 0.
\end{equation}

We extend this statement by, first, allowing the local effects to be {\it relabeled}~\cite{Haapasalo2012}.
This is the procedure of shuffling the labels of the measurement effects and possibly assigning the same label to multiple outcomes with the help of classical communication.
This leads to LOCC effects of the form $\hat{Z}^{\tilde{\rm A}\tilde{\rm B}}_{xy} = \sum_{u,v} \hat{P}^{\tilde{\rm A}}_{xy|uv}\otimes\hat{Q}^{\tilde{\rm B}}_{x,y|u,v}\in\mathcal{M}_{\rm LOCC}$ which can be obtained using LOCC~\cite{Vedral97}.
Here, $x$ and $y$ are labels to be sent to Charlie conditioned on $u$ and $v$ which characterize the outcomes obtained locally and communicated between Alice and Bob.
Note that, any LOCC POVM is necessarily separable, but the converse is not true~\cite{Bennett99}.
Next, by substituting this into Eqs.~\eqref{AVpayoff} and~\eqref{PrMeasure1}, and restricting the games to extremal ones, we define
\begin{equation}\label{measure}
\wp^\bullet(\hat{\varrho}_{\rm AB}) = \max_{\mathsf{W}^{\rm e}_{\rm sq}}\max_{\hat{Z}^{\tilde{\rm A}\tilde{\rm B}}} \overline{\wp}(\hat{\varrho}_{\rm AB};\hat{Z}^{\tilde{\rm A}\tilde{\rm B}}_{11};\mathsf{G}_{\rm sq}).
\end{equation}
Consequently, we have the following entanglement criterion as our first result.

\paragraph*{Criterion 1$^\prime$.}

A quantum state $\hat{\varrho}_{\rm AB}$ is entangled if and only if $\wp^\bullet(\hat{\varrho}_{\rm AB}) > 0$.

The proof is given in the Supplemental Material~\cite{Supp}.
The importance of Criterion~1$^\prime$ is that, in contrast to Criterion~1, it reduces the entanglement detection down to a much smaller set of games while relaxing to general LOCC measurements.

%======================================================
%======================================================
%				MDI quantification of Entanglement

\paragraph*{MDI Quantification of Entanglement.---}

It is interesting that Criterion~1$^\prime$ also provides an equivalent way to define the set of separable states as the set of all quantum states providing a maximum pay-off of zero: $\mathcal{S}_{\rm sep} = \{\hat{\varrho}| \wp^\bullet(\hat{\varrho})=0 \}$.
This also induces the idea that there exists the following continuous hierarchy of sets.

\begin{figure}[h]
  \includegraphics[width=\columnwidth]{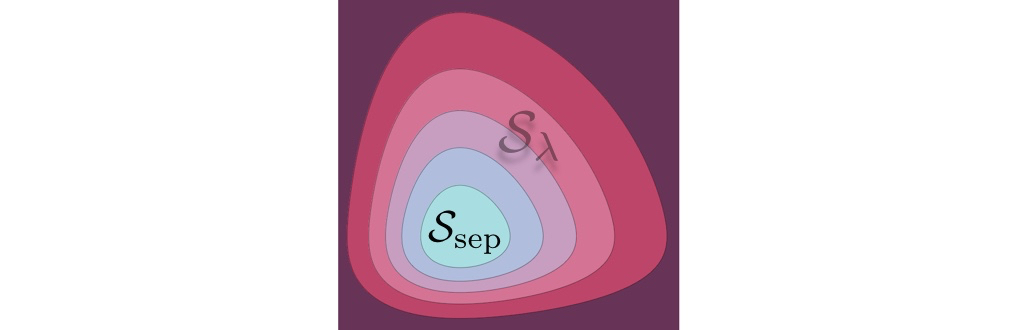}
  \caption{(Color online) The schematic representation of the continuum of the sets $\mathcal{S}_\lambda$.
  The color gradient represents the increase of the function $\wp^\bullet(\hat{\varrho}_{\rm AB})$.
  }\label{hierarchy}
\end{figure}

\paragraph*{Definition 1.}
For any $\lambda \geqslant 0$, we define $\mathcal{S}_{\lambda} = \{\hat{\varrho}_{\rm AB}| \wp^\bullet(\hat{\varrho}_{\rm AB})\leqslant \lambda \}$.
Then, for any $\lambda\geqslant 0$, $\mathcal{S}_{\rm sep} \subseteq \mathcal{S}_{\lambda}$ with $\mathcal{S}_{\rm sep} = \mathcal{S}_{\lambda}$ if and only if $\lambda = 0$; cf. Fig.~\ref{hierarchy}.

Importantly, the set $\mathcal{S}_{\lambda}$ is convex as shown in the Supplemental Material~\cite{Supp}.
Clearly, a quantum state $\hat{\varrho}_{\rm AB}$ for which $\wp^\bullet(\hat{\varrho}_{\rm AB}) > \lambda$ does not belong to the set $\mathcal{S}_{\lambda}$.
Conversely, for any $\hat{\varrho}_{\rm AB} \notin \mathcal{S}_{\lambda}$ there exists an ESQWG  $\mathsf{W}^{\rm e}_{\rm sq}\in \mathcal{W}^{\rm e}_{\rm sq}$ and an effect $\hat{Z}^{\tilde{\rm A}\tilde{\rm B}}_{11}\in\mathcal{M}_{\rm LOCC}$ for Alice and Bob such that they can obtain a pay-off value $\wp^\bullet(\hat{\varrho}_{\rm AB}) > \lambda$. 
To show this, we note that by Hahn-Banach theorem there exists a (nonextremal) witness $\hat{W}$ for the convex set $\mathcal{S}_\lambda$ which detects $\hat{\varrho}_{\rm AB}$, and that it can be optimized~\cite{Lewen2000,Sperling2009}.
The resulting optimal witness can be written as a convex combination of extremal points for which at least one of them detects $\hat{\varrho}_{\rm AB}$.

\paragraph*{Observation 1.}

$\wp^\bullet(\hat{\varrho}_{\rm AB})>\lambda$ if and only if $\hat{\varrho}_{\rm AB}\notin \mathcal{S}_{\lambda}$.

Observation~1 tells us that ESQWGs are also necessary and sufficient for characterizing the continuum of the convex sets $\mathcal{S}_{\lambda}$ via $\wp^\bullet$.
Moreover, we see that the average reward function provides a lower bound on the amount of entanglement shared by Alice and Bob.
If, for a given quantum state $\hat{\varrho}_{\rm AB}$, the reward value that Alice and Bob obtain in an ESQWG is $\overline{\wp}(\hat{\varrho}_{\rm AB};\hat{Z}^{\tilde{\rm A}\tilde{\rm B}}_{11};\mathsf{W}^{\rm e}_{\rm sq})=\lambda_0$, then $\hat{\varrho}_{\rm AB}\notin \mathcal{S}_{\lambda}$ for any $\lambda < \lambda_0$.
We now state the second result of this Letter in the theorem below and point the interested reader to the Supplemental Material for the proof~\cite{Supp}.
\paragraph*{Theorem 1.}
The pay-off $\wp^\bullet$ is a universal and faithful measure of entanglement.

By universality, we mean the value of the measure is invariant under all local invertible operations~\cite{Sperling2011PScr}.
Importantly, not only $\wp^\bullet$ is a measure of entanglement for the shared state, but also allowing the players to access infinite rounds of LOCC on input questions will not improve their best achievement.
Consequently, we can relax the LOSR restriction in ESQWGs to LOCC~\cite{Rosset2013,foot2}.
Nevertheless, it is clear that this task of measuring $\wp^\bullet$ is practically challenging in high dimensions.
Shortly, we will provide a particularly interesting scenario where the referee is only interested in the amount of NPT entanglement which, in turn, eliminates the need for the maximization over all EEWs, thus removes the aforementioned difficulty.

At this point, it is important to mention that the two maximizations in Eq.~\eqref{measure} are, in principle, performed independently by the referee and the players: Alice and Bob are responsible for the optimization of their measurements, and Charlie must choose the optimal game.
The former is less important, because $\overline{\wp}\leqslant\wp^\bullet$ guarantees that the average reward always gives a lower bound on the amount of entanglement of $\hat{\varrho}_{\rm AB}$.
To Charlie, the average reward $\overline{\wp}$ can be considered as the {\it effective entanglement} shared by Alice and Bob.
This is the amount of entanglement contained within their shared state $\hat{\varrho}_{\rm AB}$ extracted by their LOCC effect.
We note that, Charlie's payment is based on the quantum questions he prepares himself, and the coincidence statistics of the responses from Alice and Bob.
Thus, he does not need to make any assumptions about Alice and Bob's measurements in any form, as long as they are spatially separated.
However, he should hide the indices of the questions by ensuring that his questions cannot be unambiguously discriminated and that there are no side channels from his lab to Alice and Bob~\cite{Rosset2013}.
The players can increase $\overline{\wp}$ by either sharing a more entangled state or using a better LOCC strategy.
We also emphasize that there is no need for the referee to trust the players;
if the players do not perform their optimization appropriately they will incur losses.

The latter maximization requires that Charlie perform his optimization to choose the best witness (see, e.g., Ref.~\cite{Brandao2005}) before the game starts, because he must fix his choice of the game and come to an agreement with the players on it.
We also point out that it is still possible for Charlie to carry out the optimization after finishing his questions merely based on the outcome statistics.
For this, Alice and Bob chose an optimal EEW, say $\hat{W}_{{\rm e}0}$, depending on their shared state and take an optimal strategy for that.
Since the set of games that Charlie is optimizing over contains $\hat{W}_{{\rm e}0}$, then he will definitely find the pay-off expected by Alice and Bob through his optimization.
However, this requires Charlie's questions to be informationally complete so that any EEW can be expanded in terms of them~\cite{Verbanis2016,Yuan2016}.

%======================================================
%======================================================
%				MDI quantification of NPT Entanglement

\paragraph*{MDI Quantification of NPT Entanglement.---}

It is a well-known fact that there are two types of entangled states, namely, positive- and negative-partial-transpose (P- and NPT) entangled states which possess legitimate or unphysical density operators upon partial transposition operation, respectively.
It is also known that the latter is necessary for distillability~\cite{Horodecki99}, representing the importance of NPT entangled states. 
Similarly, EEWs are divided into indecomposable and decomposable classes where the latter only detects NPT entangled states~\cite{Woronowicz76,Lewen2000}.
Denoting the corresponding games as $\mathcal{W}^{\rm ie}_{\rm sq}$ and $\mathcal{W}^{\rm de}_{\rm sq}$, respectively, we have $\mathcal{W}^{\rm e}_{\rm sq}=\mathcal{W}^{\rm ie}_{\rm sq}\cup\mathcal{W}^{\rm de}_{\rm sq}$.

Now, we notice that decomposable EEWs are sufficient for detection of NPT entangled states and possess a very simple structure~\cite{Chruscinski2014}; they are of the form $\hat{W}_{\rm de}=-D|\psi\rangle\langle\psi|^{\mathsf{T}_{\rm B}}$, where $|\psi\rangle$ is a normalized entangled vector and $\mathsf{T}_{\rm B}$ denotes the partial transposition operation with respect to the second party.
In a $D\times D$ dimensional Hilbert space, we may further restrict the vectors $|\psi\rangle$ to have a Schmidt rank of $D$.
The main point here is that, for any Schmidt-rank-$D$ pure entangled state $|\psi\rangle$ there exists a stochastic LOCC (SLOCC) procedure that converts $|\psi\rangle$ into an arbitrary pure entangled state $|\phi\rangle$ with the same Schmidt rank~\cite{Nielsen,Nielsen99,Vidal99}.
The class SLOCC is a subset of separable operations, implying that the partial transpose of a SLOCC map is also SLOCC.
Therefore, $\hat{W}_{\rm de}$ can be transformed into any other decomposable EEW $\hat{V}_{\rm de}=-D|\phi\rangle\langle\phi|^{\mathsf{T}_{\rm B}}$ with a nonzero probability $0< q\leqslant 1$ via SLOCC.
In the Supplemental Material~\cite{Supp}, we discuss and prove in detail how Alice and Bob can exploit this possibility
to win a positive pay-off in an arbitrary Schmidt-rank-$D$ decomposable ESQWG chosen by Charlie, if and only if they share a NPT entangled state.
As a result, the maximization on ESQWGs in Eq.~\eqref{measure} becomes unnecessary for Charlie, if we restrict the games to Schmidt-rank-$D$ decomposable extremal ones.
It is also important to note that this restriction replaces the set $\mathcal{S}_{\rm sep}$ with the set of PPT states $\mathcal{S}_{\rm PPT}$.

\paragraph*{Theorem 2.}
For every Schmidt-rank-$D$ decomposable ESQWG $\mathsf{W}^{\rm de}_{\rm sq}$, the pay-off $\wp^\circ(\hat{\varrho}_{\rm AB};\mathsf{W}^{\rm de}_{\rm sq})=\max_{\hat{Z}^{\tilde{\rm A}\tilde{\rm B}}}\overline{\wp}(\hat{\varrho}_{\rm AB};\hat{Z}^{\tilde{\rm A}\tilde{\rm B}}_{11};\mathsf{W}^{\rm de}_{\rm sq})$ defines a universal measure of NPT entanglement.

Clearly, $\wp^\circ(\hat{\varrho}_{\rm AB};\mathsf{W}^{\rm de}_{\rm sq})\leqslant\wp^\bullet(\hat{\varrho}_{\rm AB})$ and $\wp^\circ(\hat{\varrho}_{\rm AB};\mathsf{W}^{\rm de}_{\rm sq})=0$ if and only if $\hat{\varrho}_{\rm AB}\in\mathcal{S}_{\rm PPT}$.
We also note that the universality follows from the fact that all completely positive local operations (in particular, invertible ones) preserve PPT and NPT entanglement.
Furthermore, changing the game to a different Schmidt-rank-$D$ decomposable ESQWG provides a different universal measure of NPT entanglement. 
We emphasize that, the whole procedure described here is MDI and thus, it is not possible for the players to cheat and convince the referee that they have more NPT entanglement than that contained in their state.
We also emphasize that there is no PPT entanglement for systems with dimensions up to $6$, and thus their entanglement can be perfectly characterized using $\wp^\circ$.

Consider, as an example, the Schmidt-rank-$2$ decomposable EEW $\hat{W}_{\rm de}=-2|\Psi^{-}\rangle\langle\Psi^{-}|^{\mathsf{T}_{\rm B}}$, where $|\Psi^{-}\rangle=\frac{1}{\sqrt{2}}(|01\rangle - |10\rangle)$ is a Bell state.
In a standard witnessing procedure, the Bell state $|\Phi^{+}\rangle_{\rm AB}=\frac{1}{\sqrt{2}}(|00\rangle_{\rm AB} + |11\rangle_{\rm AB})$ is detected by maximally violating the witnessing inequality of Eq.~\eqref{Wineq}, ${\rm Tr}\hat{W}_{\rm de}|\Phi^{+}\rangle_{\rm AB}\langle\Phi^{+}|=1$, while the other Bell states cannot be detected using $\hat{W}_{\rm de}$ and require different witnesses.
In an ESQWG corresponding to $\hat{W}_{\rm de}$, on the other hand, by sharing $|\Phi^{+}\rangle_{\rm AB}$ Alice and Bob will win the pay-off $\wp^\circ(|\Phi^{+}\rangle_{\rm AB})=1$, if they perform the projection onto
$\hat{Z}^{\tilde{\rm A}\tilde{\rm B}}_{11} = |\Phi^{+}\rangle_{\tilde{\rm A}}\langle\Phi^{+}|\otimes|\Phi^{+}\rangle_{\tilde{\rm B}}\langle\Phi^{+}| +
|\Phi^{-}\rangle_{\tilde{\rm A}}\langle\Phi^{-}|\otimes|\Phi^{-}\rangle_{\tilde{\rm B}}\langle\Phi^{-}| +
|\Psi^{+}\rangle_{\tilde{\rm A}}\langle\Psi^{+}|\otimes |\Psi^{+}\rangle_{\tilde{\rm B}}\langle\Psi^{+}| +
|\Psi^{-}\rangle_{\tilde{\rm A}}\langle\Psi^{-}|\otimes |\Psi^{-}\rangle_{\tilde{\rm B}}\langle\Psi^{-}|$.
Now, one would naively expect that the players could not gain a positive reward in the same game if they share instead, for instance, the state $|\Phi^{-}\rangle_{\rm AB}$, just as the witness $\hat{W}_{\rm de}$ could not detect their state in the standard witnessing procedure.
Theorem~2, however, states the contrary because the shared state is indeed NPT entangled.
It can be easily checked that if Alice and Bob project onto 
$\hat{Z}^{\tilde{\rm A}\tilde{\rm B}}_{11} = |\Phi^{-}\rangle_{\tilde{\rm A}}\langle\Phi^{-}|\otimes|\Phi^{+}\rangle_{\tilde{\rm B}}\langle\Phi^{+}| +
|\Phi^{+}\rangle_{\tilde{\rm A}}\langle\Phi^{+}|\otimes|\Phi^{-}\rangle_{\tilde{\rm B}}\langle\Phi^{-}| +
|\Psi^{-}\rangle_{\tilde{\rm A}}\langle\Psi^{-}|\otimes |\Psi^{+}\rangle_{\tilde{\rm B}}\langle\Psi^{+}| +
|\Psi^{+}\rangle_{\tilde{\rm A}}\langle\Psi^{+}|\otimes |\Psi^{-}\rangle_{\tilde{\rm B}}\langle\Psi^{-}|$,
they will obtain the pay-off $\wp^\circ(|\Phi^{-}\rangle_{\rm AB})=1$.
As a result, in accordance with Theorem~2, both $|\Phi^{+}\rangle_{\rm AB}$ and $|\Phi^{-}\rangle_{\rm AB}$ are maximally NPT entangled as measured by $\wp^\circ$.

%======================================================
%======================================================
%				Multipartite

\paragraph*{Multipartite Extension.---}

It is straightforward to extend our approach to quantify the entanglement within any partitioning of a multipartite quantum state.
In such scenarios, there are $K$ players denoted by the index set $\mathbf{I}=\{1,2,\dots,K\}$.
A $k$-partition of them is uniquely specified by the set $\mathbf{P}_k=\{\mathbf{I}_1,\dots,\mathbf{I}_k\}$ such that $\cup_{j=1}^k \mathbf{I}_j=\mathbf{I}$, and corresponds to the cut $(i\in\mathbf{I}_1|\cdots|i\in\mathbf{I}_k)$.
This time, the referee and the players first agree on a partitioning $\mathbf{P}_k$, meaning that the players within the same party $\mathbf{I}_j$ ($j=1,\dots,k$) can perform joint (global) measurements on their respective questions, while the group of players in different parties are confined to LOCC, named as $\mathbf{P}_{k}$-LOCC here.
The question set for each player is $\{\hat{\tau}^{(r)}_i\}$ ($r=1,\dots,K$), chosen at random with joint probability $\{p_i\}$, and the reward function is $\wp(x^{(1)},\dots,x^{(K)}|i)$, where $\{x^{(r)}\}$ is the set of possible answers returned by the $r$th player.
The aim of the players is to maximally win the game.

According to Refs.~\cite{Sperling2013,Shahandeh2014}, in general, multipartite entanglement has a highly complex structure.
However, the subset of witnesses extremal to the set of $\mathbf{P}_k$-separable quantum states is necessary and sufficient for detecting entanglement within $\mathbf{P}_k$.
Depending on the partitioning, Charlie thus performs the optimization over all such games denoted as $\mathcal{W}^{\mathbf{P}_k}_{\rm sq}$.

\paragraph*{Theorem~3.}
The pay-off 
\begin{equation}\label{MPmeasure}
\wp^\bullet(\hat{\varrho}_{\mathbf{P}_k}) = \max_{\mathsf{W}^{\mathbf{P}_k}_{\rm sq}\in\mathcal{W}^{\mathbf{P}_k}_{\rm sq}}\max_{\hat{Z}^{\mathbf{P}_k}\in\mathcal{M}^{\mathbf{P}_k}_{\rm LOCC}} \overline{\wp}(\hat{\varrho}_{\rm AB};\hat{Z}^{\mathbf{P}_k}_{11};\mathsf{W}^{\mathbf{P}_k}_{\rm sq})
\end{equation}
is a faithful universal measure of entanglement with respect to the partitioning $\mathbf{P}_k$.

The proof follows from the same line of proof of Theorem~1.

%======================================================
%======================================================
%				Conclusions

\paragraph*{Conclusions.---}

We showed that entanglement can be quantified operationally in a measurement-device-independent way within the context of extremal semiquantum witnessing games, a subclass of semiquantum nonlocal games, and in the LOCC paradigm.
Thus, we reduced the whole set of games down to a much smaller subset of games.
We proved that the LOCC does not help the players to increase their maximum reward for a fixed amount of effective shared entanglement.
In this way, the average reward provides a lower bound on the amount of entanglement within the shared state while the pay-off value provides a universal convex measure of entanglement.
We also showed that an arbitrary decomposable member of this class of games is necessary and sufficient for both detection and quantification of NPT entanglement, and thus, we reduced the whole set of games down to a single arbitrary game in such scenarios.
We also extended our approach to the multipartite scenario where quantification of entanglement within an arbitrary partitioning of a multipartite quantum state is desired.

%======================================================
%======================================================
%				Acknowledgements

\paragraph*{Acknowledgements.--}
The authors acknowledge useful discussions with Maciej Lewenstein and Fabio Costa.
This project was supported by the Australian Research Council Centre of Excellence for Quantum Computation and Communication Technology (CE110001027).

\appendix
\begin{widetext}

\section*{Supplemental Material: \\
Measurement-Device-Independent Approach to Entanglement Measures}

\subsection{Proof of Equation~\eqref{maxpayoffSep}}

Note that, in general, $\wp^\star(\mathcal{S}_{\rm sep};\mathsf{W}^{\rm e}_{\rm sq})\leqslant 0$.
Now, the maximization for $\wp^\star(\mathcal{S}_{\rm sep};\mathsf{W}^{\rm e}_{\rm sq})$ in Eq.~\eqref{SQWG} can be obtained by choosing the POVMs $\hat{P}^{\tilde{\rm A}}_1$ and $\hat{Q}^{\tilde{\rm B}}_1$ to be the projections onto $|\Phi^{+}\rangle=d^{-\frac{1}{2}}\sum_i |i,i\rangle$, and $\hat{\varrho}_{\rm AB}$ to be the transpose of the extremal point of $\hat{W}_{\rm e}$, $\hat{\sigma}^{\mathsf{T}}_{\rm AB}=\hat{\sigma}_{\rm AB}=|a,b\rangle_{\rm AB}\langle a,b|$, which gives $\wp^\star(\hat{\sigma}_{\rm AB};\mathsf{W}^{\rm e}_{\rm sq})=d^{-2}{\rm Tr}\hat{W}\hat{\sigma}_{\rm AB} = 0$.
Thus, by making use of Eq.~\eqref{SepSQNLequal}, we obtain $\wp^\star(\mathcal{S}_{\rm sep};\mathsf{W}^{\rm e}_{\rm sq}) = 0$.\qed

\subsection{Proof of Citerion~1$^\prime$}

It is sufficient to prove that a quantum state $\hat{\sigma}\in\mathcal{S}_{\rm sep}$ if and only if $\wp^\bullet(\hat{\sigma}_{\rm AB})=0$.
The proof is as follows.
For any state $\hat{\varrho}_{\rm AB}$, we can write
\begin{equation}\label{C1Proof}
\begin{split}
\wp^\bullet(\hat{\varrho}_{\rm AB})&=\max_{\hat{W}_e}
\max_{\hat{Z}^{\tilde{\rm A}\tilde{\rm B}}\in\mathcal{M}_{{\rm LOCC}}} {\rm Tr}\hat{Z}^{\tilde{\rm A}\tilde{\rm B}}_{11}(\hat{W}_{\rm e}\otimes\hat{\varrho}_{\rm AB})\\
&=\max_{\hat{W}_e}\max_{\hat{Z}^{\tilde{\rm A}\tilde{\rm B}}\in\mathcal{M}_{{\rm LOCC}}} {\rm Tr}(\sum_{u,v} \hat{P}^{\tilde{\rm A}}_{11|uv}\otimes\hat{Q}^{\tilde{\rm B}}_{11|u,v})(\hat{W}_{\rm e}\otimes\hat{\varrho}_{\rm AB})\\
& \leqslant\max_{\hat{W}_e}\sum_{u,v}\max_{\hat{P}^{\tilde{\rm A}}\in\mathcal{M}_{\tilde{\rm A}},\hat{Q}^{\tilde{\rm B}}\in\mathcal{M}_{\tilde{\rm B}}} {\rm Tr}( \hat{P}^{\tilde{\rm A}}_{11|uv}\otimes\hat{Q}^{\tilde{\rm B}}_{11|u,v})(\hat{W}_{\rm e}\otimes\hat{\varrho}_{\rm AB})\\
&=\max_{\hat{W}_e}\sum_{u,v}\wp^\star(\hat{\varrho}_{\rm AB};\mathsf{W}^{\rm e}_{\rm sq}).
\end{split}
\end{equation}
To prove the necessary part, we note that for every separable state $\hat{\sigma}_{\rm AB}\in\mathcal{S}_{\rm sep}$, the right-hand-side of Eq.~\eqref{C1Proof} is just zero by using Eq.~\eqref{maxpayoffSep}.
Thus, the pay-off can never exceed zero, and hence, the players can at best obtain the same pay-off as $\wp^\star(\mathcal{S}_{\rm sep};\mathsf{W}^{\rm e}_{\rm sq})=0$.

The sufficient part follows from the fact that, using Criteria~1, for every entangled state $\hat{\varrho}_{\rm AB}\notin\mathcal{S}_{\rm sep}$ there exists an extremal witness $\hat{W}_{\rm e}$ such that $\wp^\star(\hat{\varrho}_{\rm AB};\mathsf{W}^{\rm e}_{\rm sq})>0$.
Since all the summands on the right-hand-side of Eq.~\eqref{C1Proof} are nonnegative, it follows that $\wp^\bullet(\hat{\varrho}_{\rm AB})>0$.

\subsection{Convexity and inclusions of the sets $\mathcal{S}_\lambda$}

The set $\mathcal{S}_\lambda$ being convex means that
\begin{equation}
\forall \hat{\varrho}_{\rm AB},\hat{\sigma}_{\rm AB} \in \mathcal{S}_{\lambda}  \text{~and~} p\in [0,1], \quad \hat{\eta}_{\rm AB}= p \hat{\varrho}_{\rm AB} + (1-p) \hat{\sigma}_{\rm AB} \in \mathcal{S}_\lambda.
\end{equation}
It is sufficient to show that
\begin{equation}
\wp^\bullet(\hat{\eta}_{\rm AB})\leqslant p \wp^\bullet(\hat{\varrho}_{\rm AB}) + (1-p) \wp^\bullet(\hat{\sigma}_{\rm AB}).
\end{equation}
We note that
\begin{equation}
\begin{split}
\wp^\bullet(\hat{\eta}_{\rm AB}) &= \max_{\hat{W}_{\rm e}} \max_{\hat{Z}^{\tilde{\rm A}\tilde{\rm B}}\in\mathcal{M}_{{\rm LOCC}}} {\rm Tr}\hat{Z}^{\tilde{\rm A}\tilde{\rm B}}_1(\hat{W}_{\rm e}\otimes\hat{\eta}_{\rm AB})\\
&= \max_{\hat{W}_{\rm e}} \max_{\hat{Z}^{\tilde{\rm A}\tilde{\rm B}}\in\mathcal{M}_{{\rm LOCC}}} \left\{ p {\rm Tr}\hat{Z}^{\tilde{\rm A}\tilde{\rm B}}_1(\hat{W}_{\rm e}\otimes\hat{\varrho}_{\rm AB}) + (1-p) {\rm Tr}\hat{Z}^{\tilde{\rm A}\tilde{\rm B}}_1(\hat{W}_{\rm e}\otimes\hat{\sigma}_{\rm AB})\right\}\\
& \leqslant p \max_{\hat{W}_{\rm e}}\max_{\hat{Z}^{\tilde{\rm A}\tilde{\rm B}}\in\mathcal{M}_{{\rm LOCC}}} {\rm Tr}\hat{Z}^{\tilde{\rm A}\tilde{\rm B}}_1(\hat{W}_{\rm e}\otimes\hat{\varrho}_{\rm AB}) + (1-p) \max_{\hat{W}_{\rm e}}\max_{\hat{Z}^{\tilde{\rm A}\tilde{\rm B}}\in\mathcal{M}_{{\rm LOCC}}} {\rm Tr}\hat{Z}^{\tilde{\rm A}\tilde{\rm B}}_1(\hat{W}_{\rm e}\otimes\hat{\sigma}_{\rm AB})\\
&= p \wp^\bullet(\hat{\varrho}_{\rm AB}) + (1-p) \wp^\bullet(\hat{\sigma}_{\rm AB}),
\end{split}
\end{equation}
where the third inequality is a result of the fact that $\max_x \left\{ p f(x,y_1) + (1-p) f(x,y_2) \right\} \leqslant p \max_x f(x,y_1) + (1-p) \max_x f(x,y_2)$.

The inclusions are trivial.
Also, the case of $\lambda=0$ is already proven in Criteria~1$^\prime$.

\subsection{Proof of Theorem~1} 

We need to prove that, (i) $\wp^\bullet(\hat{\varrho}_{\rm AB})=0$ if and only if $\hat{\varrho}_{\rm AB}\in \mathcal{S}_{\rm sep}$, and, (ii) $\wp^\bullet(\hat{\varrho}_{\rm AB}) \geqslant \wp^\bullet(\hat{\varrho}_{\rm AB}')$, where $\hat{\varrho}_{\rm AB}'=\Lambda(\hat{\varrho}_{\rm AB})/{\rm Tr}\Lambda(\hat{\varrho}_{\rm AB})$ for any LOCC operation $\Lambda\in\mathcal{C}_{\rm LOCC}$.

The first condition is already proven as Criterion~1$^\prime$.
Let us prove the condition (ii) in a more general context, by allowing the players to make separable operations, $\Lambda\in\mathcal{C}_{\rm sep}$, on both their respective parts of shared state and quantum questions.
Given that Alice and Bob receive the ensemble of questions $\hat{\varpi}_{{\rm A}_0{\rm B}_0} = \sum_i p_i \hat{\tau}^{{\rm A}_0}_i \otimes \hat{\omega}^{{\rm B}_0}_i$, any normalized separable operation can be written as a convex combination of completely positive trace-preserving separable operations, with Kraus decomposition $\hat{\varpi}_{{\rm A}_0{\rm B}_0} \otimes \hat{\varrho}_{\rm AB} \mapsto \Lambda_{\rm n}(\hat{\varpi}_{{\rm A}_0{\rm B}_0} \otimes \hat{\varrho}_{\rm AB})= \sum_j q_{j}\hat{F}_{j} (\hat{\varpi}_{{\rm A}_0{\rm B}_0} \otimes \hat{\varrho}_{\rm AB}) \hat{F}^\dag_{j}$ where $\sum_j q_{j} = 1$ and $\hat{F}_j=\hat{F}^{\tilde{\rm A}}_j\otimes\hat{F}^{\tilde{\rm B}}_j$.

Using the dual operation, we have that ${\rm Tr}\hat{Z}^{\tilde{\rm A}\tilde{\rm B}}_{11} \Lambda_{\rm n}(\hat{W}_{\rm e} \otimes \hat{\varrho}_{\rm AB}) = {\rm Tr}\hat{X}^{\tilde{\rm A}\tilde{\rm B}}_{11} (\hat{W}_{\rm e} \otimes \hat{\varrho}_{\rm AB})$, in which $\hat{X}^{\tilde{\rm A}\tilde{\rm B}}_{11} = \Lambda_{\rm n}^\dag (\hat{Z}^{\tilde{\rm A}\tilde{\rm B}}_{11}) = \sum_j q_{j}\hat{F}^\dag_{j} \hat{Z}^{\tilde{\rm A}\tilde{\rm B}}_{11} \hat{F}_{j}$ is a convex combination of LOCC effects for Alice and Bob.
We note that any convex combination of LOCC effects is also a LOCC effect.
Therefore, it holds true that
\begin{equation}\label{nonincreasproof}
\max_{\hat{X}^{\tilde{\rm A}\tilde{\rm B}}\in\mathcal{M}_{{\rm LOCC}}} {\rm Tr}\hat{X}^{\tilde{\rm A}\tilde{\rm B}}_{11} (\hat{W}_{\rm e} \otimes \hat{\varrho}_{\rm AB})\leqslant \max_{\hat{Z}^{\tilde{\rm A}\tilde{\rm B}}\in\mathcal{M}_{{\rm LOCC}}} {\rm Tr}\hat{Z}^{\tilde{\rm A}\tilde{\rm B}}_{11} (\hat{W}_{\rm e} \otimes \hat{\varrho}_{\rm AB}).
\end{equation}
The inequality is obtained by noticing that a linear function on a convex set achieves its maximum at the extremal points.
We note that the particular requirement of $\wp^\bullet(\hat{\varrho}'_{\rm AB})\leqslant \wp^\bullet(\hat{\varrho}_{\rm AB})$ follows as a special case, after maximizing both sides of Eq.~\eqref{nonincreasproof} over all EEWs, if no operation is done on the input questions.

Finally, a measure of entanglement is universal if it is invariant under all local invertible operations.
This is shown to be the case if the measure is nonincreasing under all separable operations~\cite{S2011PScr}, which holds true for $\wp^\bullet$. \qed

\subsection{Proof of Theorem~2} 

Very similar to the proof of Theorem~1, we need to prove that, (i) $\wp^\circ(\hat{\varrho}_{\rm AB};\mathsf{W}^{\rm de}_{\rm sq})=0$ if and only if $\hat{\varrho}_{\rm AB}\in \mathcal{S}_{\rm PPT}$, and, (ii) $\wp^\circ(\hat{\varrho}_{\rm AB};\mathsf{W}^{\rm de}_{\rm sq}) \geqslant \wp^\circ(\hat{\varrho}_{\rm AB}';\mathsf{W}^{\rm de}_{\rm sq})$, where $\hat{\varrho}_{\rm AB}'=\Lambda(\hat{\varrho}_{\rm AB})/{\rm Tr}\Lambda(\hat{\varrho}_{\rm AB})$ for any LOCC operation $\Lambda\in\mathcal{C}_{\rm LOCC}$.

First, let us prove that Alice and Bob are able to win a positive pay-off by sharing any NPT entangled state in any arbitrary decomposable ESQWG played by the referee.
 
Suppose that there are two witnesses $\hat{W}_{\rm de}=-D|\psi\rangle\langle\psi|^{\mathsf{T}_{\rm B}}=\sum_i \beta_i \hat{\tau}^{{\rm A}_0\mathsf{T}}_i \otimes \hat{\omega}^{{\rm B}_0\mathsf{T}}_i$ and $\hat{V}_{\rm de}=-D|\phi\rangle\langle\phi|^{\mathsf{T}_{\rm B}}$.
The former, corresponding to the game played by Charlie, does not detect the NPT entangled state $\hat{\varrho}_{\rm AB}$ shared by Alice and Bob in the standard witnessing scenario.
The latter, however, detects it.
We know that $|\psi\rangle$ with Schmidt-rank $D$ can be transformed into $|\phi\rangle$ with Schmidt-rank $R\leqslant D$ via a stochastic LOCC (SLOCC) map $\Upsilon$ with the success probability $0<q\leqslant1$~\cite{N,N99,V99}.
That is $q^{-1}\Upsilon(|\psi\rangle\langle\psi|)=|\phi\rangle\langle\phi|$, where $q={\rm Tr}\Upsilon(|\psi\rangle\langle\psi|)$. 
SLOCC operations are a subset of separable ones, and thus linear, with Kraus decompositions of the form $\Upsilon(\cdot)=\sum_j \hat{A}_j\otimes\hat{B}_j(\cdot)\hat{A}^\dag_j\otimes\hat{B}^\dag_j$.
Thus, their partial transpose, $\Upsilon^{\mathsf{T}_{\rm B}}(\cdot)=\sum_j \hat{A}_j\otimes\hat{B}^*_j(\cdot)\hat{A}^\dag_j\otimes\hat{B}^{\mathsf{T}}_j$, is completely positive, and in fact, SLOCC.
Therefore, $q^{-1}\Upsilon^{\mathsf{T}_{\rm B}}(\hat{W}_{\rm de})=\hat{V}_{\rm de}$.
This means that any decomposable EEW with Schmidt rank $D$ can be transformed into an arbitrary decomposable EEW with Schmidt rank $R\leqslant D$ with nonzero probability $q$.
It is clear that the players do not have direct access to the witness as the reward function is fixed by the referee to correspond to $\hat{W}_{\rm de}$ --- Alice and Bob cannot negotiate with the referee on that --- however, they have the possibility to operate with $\Upsilon^{\mathsf{T}_{\rm B}}$ on their respective questions.
Then, if Alice and Bob can win a game $\mathsf{V}^{\rm de}_{\rm sq}$ corresponding to witness $\hat{V}_{\rm de}$, i.e., achieve an average reward $\overline{\wp}(\hat\varrho_{\rm AB};\hat{Z}^{\tilde{\rm A} \tilde{\rm B}}_{11};\mathsf{V}^{\rm de}_{\rm sq})>0$ via measuring some joint LOCC effect $\hat{Z}^{\tilde{\rm A} \tilde{\rm B}}_{11}$, we have
\begin{equation}\label{detW}
\begin{split}
 \overline{\wp}(\hat{\varrho}_{\rm AB};\hat{Z}^{\tilde{\rm A} \tilde{\rm B}}_{11};\mathsf{W}^{\rm de}_{\rm sq}|\Upsilon^{\mathsf{T}_{\rm B}}) & = \sum_i \beta_i {\rm Tr}\hat{Z}^{\tilde{\rm A}\tilde{\rm B}}_{11} (\Upsilon^{\mathsf{T}_{\rm B}}(\hat{\tau}^{{\rm A}_0\mathsf{T}}_i \otimes \hat{\omega}^{{\rm B}_0\mathsf{T}}_i)\otimes\hat{\varrho}_{\rm AB})\\
& = {\rm Tr}\hat{Z}^{\tilde{\rm A}\tilde{\rm B}}_{11} (\Upsilon^{\mathsf{T}_{\rm B}}(\hat{W}_{\rm de})\otimes\hat{\varrho}_{\rm AB})\\
& = q {\rm Tr}\hat{Z}^{\tilde{\rm A}\tilde{\rm B}}_{11}(\hat{V}_{\rm de}\otimes\hat{\varrho}_{\rm AB})\\
& = q \overline{\wp}(\hat{\varrho}_{\rm AB};\hat{Z}^{\tilde{\rm A} \tilde{\rm B}}_{11};\mathsf{V}^{\rm de}_{\rm sq})>0.
\end{split}
\end{equation}
The result is positive because $\hat{V}_{\rm de}$ detects the entanglement of $\hat{\varrho}_{\rm AB}$ upon measurement of $\hat{Z}^{\tilde{\rm A}\tilde{\rm B}}_{11}$, no matter how small the probability $q$ is. 
Equation~\eqref{detW} simply implies that the players can always win a positive pay-off (even though very small due to the probability $q$ being small) by making use of an appropriate SLOCC strategy $\Upsilon^{\mathsf{T}_{\rm B}}$, if their shared state is NPT entangled, regardless of the decomposable EEW $\hat{W}_{\rm de}$ chosen by the referee.

Importantly, we may also rearrange Eq.~\eqref{detW} as
\begin{equation}\label{detX}
\begin{split}
q \overline{\wp}(\hat{\varrho}_{\rm AB};\hat{Z}^{\tilde{\rm A}\tilde{\rm B}}_{11};\mathsf{V}^{\rm de}_{\rm sq}) & = {\rm Tr}\hat{Z}^{\tilde{\rm A}\tilde{\rm B}}_{11} (\Upsilon^{\mathsf{T}_{\rm B}}(\hat{W}_{\rm de})\otimes\hat{\varrho}_{\rm AB})\\
 & = {\rm Tr}\Upsilon^{\dag\mathsf{T}_{\rm B}}(\hat{Z}^{\tilde{\rm A}\tilde{\rm B}}_{11}) (\hat{W}_{\rm de}\otimes\hat{\varrho}_{\rm AB})\\
& = {\rm Tr}\hat{X}^{\tilde{\rm A}\tilde{\rm B}}_{11}(\hat{W}_{\rm de}\otimes\hat{\varrho}_{\rm AB})\\
& = \overline{\wp}(\hat{\varrho}_{\rm AB};\hat{X}^{\tilde{\rm A} \tilde{\rm B}}_{11};\mathsf{W}^{\rm de}_{\rm sq}).
\end{split}
\end{equation}
Equation~\eqref{detX} represents the possibility of considering SLOCC as part of the LOCC measurement strategy, $\hat{X}^{\tilde{\rm A}\tilde{\rm B}}_{11}=\Upsilon^{\dag\mathsf{T}_{\rm B}}(\hat{Z}^{\tilde{\rm A}\tilde{\rm B}}_{11})$, which incorporates the probability of success of the operation as part of the probability of obtaining the $11$ outcome corresponding to the effect $\hat{X}^{\tilde{\rm A}\tilde{\rm B}}_{11}$.
That is, ${\rm Pr}(\hat{W}_{\rm de}\rightarrow \hat{V}_{\rm de})\times {\rm Pr}(\hat{Z}^{\tilde{\rm A}\tilde{\rm B}}_{11}|\hat{V}_{\rm de}\otimes\hat{\varrho}_{\rm AB})={\rm Pr}(\hat{X}^{\tilde{\rm A}\tilde{\rm B}}_{11}|\hat{W}_{\rm de}\otimes\hat{\varrho}_{\rm AB})$.

As a result of the above discussion, condition (i) follows immediately from the fact that for any NPT entangled state in a $D\times D$ dimensional Hilbert space, there exists a decomposable EEW $\hat{V}_{\rm de}=-D|\phi\rangle\langle\phi|^{\mathsf{T}_{\rm B}}$ with the Schmidt rank of at most $D$ which detects it.
Thus, a state is PPT if and only if a detection is impossible (recall that a decomposable witness can only detect the entanglement of NPT states), which in turn implies $\wp^\circ(\hat{\varrho}_{\rm AB};\mathsf{W}^{\rm de}_{\rm sq})=0$.

Now, we prove the condition (ii) in a more general context, again, by allowing the players to make further separable operations on both their respective parts of the shared state and quantum questions.
Given that Alice and Bob receive the ensemble of questions $\hat{\varpi}_{{\rm A}_0{\rm B}_0} = \sum_i p_i \hat{\tau}^{{\rm A}_0}_i \otimes \hat{\omega}^{{\rm B}_0}_i$, any normalized separable operation can be written as a convex combination of completely positive trace-preserving separable operations, with Kraus decomposition $\hat{\varpi}_{{\rm A}_0{\rm B}_0} \otimes \hat{\varrho}_{\rm AB} \mapsto \Lambda_{\rm n}(\hat{\varpi}_{{\rm A}_0{\rm B}_0} \otimes \hat{\varrho}_{\rm AB})= \sum_j q_{j}\hat{F}_{j} (\hat{\varpi}_{{\rm A}_0{\rm B}_0} \otimes \hat{\varrho}_{\rm AB}) \hat{F}^\dag_{j}$ where $\sum_j q_{j} = 1$ and $\hat{F}_j=\hat{F}^{\tilde{\rm A}}_j\otimes\hat{F}^{\tilde{\rm B}}_j$.

Suppose that $\hat{V}_{\rm de}=-D|\phi\rangle\langle\phi|^{\mathsf{T}_{\rm B}}$ is the witness that detects the shared state $\hat{\varrho}_{\rm AB}$, while Alice and Bob play the game $\mathsf{W}^{\rm de}_{\rm sq}$ corresponding to the witness $\hat{W}_{\rm de}=-D|\psi\rangle\langle\psi|^{\mathsf{T}_{\rm B}}$ with Charlie.
The players make a SLOCC on their respective questions (or equivalently, on their effects) as described in Eq.~\eqref{detW}.
Moreover, suppose that an appropriate measurement denoted by $\hat{X}^{\tilde{\rm A}\tilde{\rm B}}_{11}$ gives the optimal pay-off for the game $\mathsf{W}^{\rm de}_{\rm sq}$. 
That is, in Eq.~\eqref{detW}, $\wp^\circ(\hat{\varrho}_{\rm AB};\mathsf{W}^{\rm de}_{\rm sq}) = q\overline{\wp}(\hat{\varrho}_{\rm AB};\hat{Z}^{\tilde{\rm A}\tilde{\rm B}}_{11};\mathsf{V}^{\rm de}_{\rm sq})$.
We prove that further LOCC operations $\Lambda_{\rm n}$ on the shared state cannot increase the pay-off.
Using the dual operation, we have that ${\rm Tr}\hat{X}^{\tilde{\rm A}\tilde{\rm B}}_{11} \Lambda_{\rm n}(\hat{W}_{\rm de} \otimes \hat{\varrho}_{\rm AB}) = {\rm Tr}\hat{Y}^{\tilde{\rm A}\tilde{\rm B}}_{11} (\hat{W}_{\rm de} \otimes \hat{\varrho}_{\rm AB})$, in which $\hat{Y}^{\tilde{\rm A}\tilde{\rm B}}_{11} = \Lambda_{\rm n}^\dag (\hat{X}^{\tilde{\rm A}\tilde{\rm B}}_{11}) = \sum_j q_{j}\hat{F}^\dag_{j} \hat{X}^{\tilde{\rm A}\tilde{\rm B}}_{11} \hat{F}_{j}$ is a convex combination of LOCC effects for Alice and Bob.
We note that any convex combination of LOCC effects is also a LOCC effect and thus, it holds true that
\begin{equation}\label{nonincreasproof}
\max_{\hat{Y}^{\tilde{\rm A}\tilde{\rm B}}\in\mathcal{M}_{{\rm LOCC}}} {\rm Tr}\hat{Y}^{\tilde{\rm A}\tilde{\rm B}}_{11} (\hat{W}_{\rm de} \otimes \hat{\varrho}_{\rm AB})\leqslant \max_{\hat{X}^{\tilde{\rm A}\tilde{\rm B}}\in\mathcal{M}_{{\rm LOCC}}} {\rm Tr}\hat{X}^{\tilde{\rm A}\tilde{\rm B}}_{11} (\hat{W}_{\rm de} \otimes \hat{\varrho}_{\rm AB}).
\end{equation}
The inequality is obtained by noticing that a linear function on a convex set achieves its maximum at the extremal points.
We note that the particular requirement of $\wp^\circ(\hat{\varrho}'_{\rm AB};\mathsf{W}^{\rm de}_{\rm sq})\leqslant \wp^\circ(\hat{\varrho}_{\rm AB};\mathsf{W}^{\rm de}_{\rm sq})$ follows as a special case, if no operation is done on the input questions.

As before, $\wp^\circ$ is invariant under all local invertible operations on the state, due to the fact that such operations are a subset of all separable operations. 
Therefore, it is a universal measure of NPT entanglement.\qed

We emphasize that for a different game $\mathsf{J}^{\rm de}_{\rm sq}$ corresponding to the Schmidt rank-$D$ decomposable EEW $\hat{J}_{\rm de}=-D|\xi\rangle\langle\xi|^{\mathsf{T}_{\rm B}}$, one get another universal measure of NPT entanglement $\wp^\circ(\hat{\varrho}_{\rm AB};\mathsf{J}^{\rm de}_{\rm sq})$. 
However, if $\hat{J}_{\rm de}$ can be obtained from $\hat{W}_{\rm de}$ using LOCC (that is, deterministically) then $\wp^\circ(\hat{\varrho}_{\rm AB};\mathsf{J}^{\rm de}_{\rm sq})=\wp^\circ(\hat{\varrho}_{\rm AB};\mathsf{W}^{\rm de}_{\rm sq})$.
It is known that $|\psi\rangle$ is LOCC convertible to $|\xi\rangle$ if and only if the vector of Schmidt coefficients of $|\psi\rangle$, $\vec{\mu}_\psi$, is majorized by the vector of Schmidt coefficients of $|\xi\rangle$, $\vec{\mu}_\xi$, denoted by $\vec{\mu}_\psi \prec \vec{\mu}_\xi$~\cite{N,NV01}.
In particular, the vector of Schmidt coefficients of $|\psi^{(D)}_{\rm Bell}\rangle = \frac{1}{\sqrt{D}}\sum_{i=1}^D |i,i\rangle$ is majorized by any vector with Schmidt rank $D$.
As a result, $|\psi^{(D)}_{\rm Bell}\rangle$ can be transform into any Schmidt rank $D$ vector using (deterministic) LOCC.
In light of the previous discussion, Charlie can choose the witness to be $\hat{W}_{\rm de}=|\psi^{(D)}_{\rm Bell}\rangle\langle\psi^{(D)}_{\rm Bell}|^{\mathsf{T}_{\rm B}}=\hat{S}$, that is the swap operator $\hat{S}$, and make the game easier for Alice and Bob, while remaining robust against cheating.

\end{widetext}

\begin{thebibliography}{100}
\bibitem{foot1} Note that our terminology here is slightly different from that of Buscemi.
As we will see shortly, the pay-off made by Charlie is always considered by him to be the best Alice and Bob could do.
\end{thebibliography}
\end{document}